# Absorption and Stimulated Emission by a Thin Slab Obeying the Lorentz Oscillator Model

Masud Mansuripur

College of Optical Sciences, The University of Arizona, Tucson, Arizona 85721 (U.S.A.)

**Abstract**. In his celebrated 1916-17 papers in which he proposed the *A* and *B* coefficients for the spontaneous and stimulated emission of energy quanta from excited atoms, Einstein conjectured that stimulated emission involves the release of individual quanta (later dubbed "photons") along the direction of an incident photon with the same energy, momentum, phase, and polarization state as that of the incident photon. According to classical electrodynamics, of course, an oscillating dipole must radiate an azimuthally symmetric electromagnetic field around its axis of oscillation. Nevertheless, Einstein suggested that the release of stored energy from excited atoms in the form of discrete quanta (photons) must be directional, and that, in the case of stimulated emission, the direction of the emitted photon must coincide with that of the incident photon. The goal of the present paper is to show that some of the prominent features of absorption and stimulated emission emerge from Maxwellian electrodynamics in conjunction with the simple mass-and-spring model of an atom known as the Lorentz oscillator model.

**1. Introduction**. The goal of the present paper is to show that some of the prominent features of absorption and stimulated emission that were originally discussed in Einstein's 1916-17 papers [1-4], emerge from the classical theory of electrodynamics in conjunction with the simple mass-and-spring model of an atom known as the Lorentz oscillator model [5-8]. Consider a thin slab of lossy/gainy medium of thickness $d$ and susceptibility $\varepsilon_0 \chi(\omega)$, where

$$\chi(\omega) = \chi_b \pm \frac{\omega_p^2}{\omega_r^2 - \omega^2 - \mathrm{i}\gamma\omega}. \tag{1}$$

Here $\varepsilon_0$ is the permittivity of free space, $\chi_b$ is the background susceptibility, $\omega_r$ is the resonance frequency, $\omega_p$ is the plasma frequency, and $\gamma$ is the damping (or amplification) coefficient of the medium. The $\pm$ signs specify whether the medium is lossy (plus sign) or gainy (minus sign). Note that $\gamma$ is always positive, and that the sign of the oscillator strength (represented by $\pm$ in Eq.(1)), rather than that of $\gamma$, determines the lossy/gainy nature of the medium. The susceptibility given by Eq.(1), and also the corresponding refractive index, thus satisfy the Kramers-Kronig relations so long as the poles of $\chi(\omega)$ remain in the lower-half of the complex $\omega$-plane.

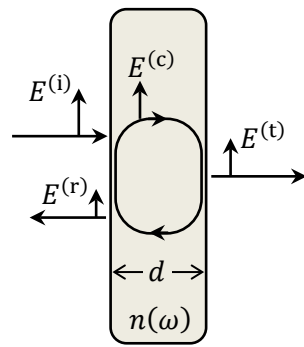

**Fig.1**. A linearly-polarized plane wave of amplitude $E^{(\mathrm{i})}$ and frequency $\omega$ is normally incident on a thin dielectric slab of thickness $d$ and refractive index $n(\omega)$. The reflected $E$-field amplitude is $E^{(\mathrm{r})}$, the transmitted field is $E^{(\mathrm{t})}$, and the amplitude of the $E$-field circulating inside the slab is $E^{(\mathrm{c})}$.

Let us denote by $E^{(\mathrm{i})}$, $E^{(\mathrm{r})}$, $E^{(\mathrm{t})}$ the $E$-field amplitudes of incident, reflected, and transmitted plane-waves at normal incidence at the frequency $\omega$, and by $E^{(\mathrm{c})}$ the amplitude of the circulating $E$-field inside the slab, as shown in Fig.1. Denoting by $n(\omega) = \sqrt{1 + \chi(\omega)}$ the refractive index of the material medium, the Fresnel reflection and transmission coefficients of the vacuum-dielectric interface at the front facet of the slab are given by [5,6]

$$\rho_{12}(\omega) = \frac{1 - n(\omega)}{1 + n(\omega)}, \tag{2a}$$

$$\tau_{12}(\omega) = \frac{2}{1 + n(\omega)}. \tag{2b}$$

The corresponding Fresnel coefficients of the dielectric-vacuum interface (i.e., from inside the slab to the outside) are

$$\rho_{21}(\omega) = \frac{n(\omega)-1}{n(\omega)+1}, \tag{3a}$$

$$\tau_{21}(\omega) = \frac{2n(\omega)}{n(\omega)+1}. \tag{3b}$$

We may now write

$$E^{(c)} = \tau_{12}E^{(i)} + \rho_{21}^2 \exp[i2n(\omega)\omega d/c]\, E^{(c)}, \tag{4}$$

$$E^{(r)} = \rho_{12}E^{(i)} + \tau_{21}\rho_{21} \exp[i2n(\omega)\omega d/c]\, E^{(c)}, \tag{5}$$

$$E^{(t)} = \tau_{21} \exp[in(\omega)\omega d/c]\, E^{(c)}. \tag{6}$$

In the above equations, $c = 1/\sqrt{\mu_0 \varepsilon_0}$ is the speed of light in vacuum, $\varepsilon_0$ and $\mu_0$ being the permittivity and permeability of free space, respectively. Upon solving Eq.(4) for $E^{(c)}$, and incorporating the result into Eqs.(5) and (6), we find

$$\frac{E^{(r)}}{E^{(i)}} = \frac{\rho_{12} + (\tau_{12}\tau_{21} - \rho_{12}\rho_{21})\rho_{21}\exp[i2n(\omega)\omega d/c]}{1 - \rho_{21}^2 \exp[i2n(\omega)\omega d/c]}, \tag{7}$$

$$\frac{E^{(t)}}{E^{(i)}} = \frac{\tau_{12}\tau_{21}\exp[in(\omega)\omega d/c]}{1 - \rho_{21}^2 \exp[i2n(\omega)\omega d/c]}. \tag{8}$$

Substitution from Eqs.(2) and (3) into Eqs.(7) and (8) now yields

$$\frac{E^{(r)}}{E^{(i)}} = \frac{[n^2(\omega)-1]\{\exp[i2n(\omega)\omega d/c] - 1\}}{[n(\omega)+1]^2 - [n(\omega)-1]^2 \exp[i2n(\omega)\omega d/c]}, \tag{9}$$

$$\frac{E^{(t)}}{E^{(i)}} = \frac{4n(\omega)\exp[in(\omega)\omega d/c]}{[n(\omega)+1]^2 - [n(\omega)-1]^2 \exp[i2n(\omega)\omega d/c]}. \tag{10}$$

Equations (9) and (10) are the exact expressions of the Fresnel reflection and transmission coefficients for a slab of thickness $d$ and refractive index $n(\omega)$, surrounded by vacuum and illuminated at normal incidence by a monochromatic plane-wave of frequency $\omega$. For a sufficiently thin slab, we will have $\exp[in(\omega)\omega d/c] \cong 1 + in(\omega)\omega d/c$, and, therefore,

$$\frac{E^{(r)}}{E^{(i)}} \cong \frac{[n^2(\omega)-1](i\omega d/c)}{2 - [n(\omega)-1]^2(i\omega d/c)} \cong \tfrac{1}{2}[n^2(\omega) - 1](i\omega d/c). \tag{11}$$

$$\frac{E^{(t)}}{E^{(i)}} \cong \frac{1 + n(\omega)(i\omega d/c)}{1 - \tfrac{1}{2}[n(\omega)-1]^2(i\omega d/c)} \cong 1 + \tfrac{1}{2}[n^2(\omega) + 1](i\omega d/c). \tag{12}$$

Note that the $E$-field radiated by the thin slab has the same amplitude in the reflection and transmission directions, with the only difference between $E^{(r)}$ and $E^{(t)}$ in Eqs.(11) and (12) being the addition of the (properly delayed) incident amplitude, namely, $E^{(i)} \exp(i\omega d/c) \cong E^{(i)}(1 + i\omega d/c)$, to the radiated beam in the transmission direction.

Upon substituting $1 + \chi(\omega)$ for $n^2(\omega)$ in Eqs.(11) and (12)—without any additional approximations—we find

$$\frac{E^{(r)}}{E^{(i)}} \cong \tfrac{1}{2}\left(\chi_b \pm \frac{\omega_p^2}{\omega_r^2 - \omega^2 - i\gamma\omega}\right)(i\omega d/c) = \mp \frac{\tfrac{1}{2}(\omega_p^2 \gamma d/c)\omega^2}{(\omega_r^2 - \omega^2)^2 + \gamma^2 \omega^2} + \tfrac{1}{2}i\left[\chi_b \pm \frac{\omega_p^2(\omega_r^2 - \omega^2)}{(\omega_r^2 - \omega^2)^2 + \gamma^2 \omega^2}\right](\omega d/c). \tag{13}$$

$$\frac{E^{(t)}}{E^{(i)}} \cong 1 + \left(1 + \tfrac{1}{2}\chi_b \pm \frac{\tfrac{1}{2}\omega_p^2}{\omega_r^2 - \omega^2 - i\gamma\omega}\right)(i\omega d/c)$$

$$= \left[1 \mp \frac{\tfrac{1}{2}(\omega_p^2 \gamma d/c)\omega^2}{(\omega_r^2 - \omega^2)^2 + \gamma^2\omega^2}\right] + i\left[1 + \tfrac{1}{2}\chi_b \pm \frac{\tfrac{1}{2}\omega_p^2(\omega_r^2 - \omega^2)}{(\omega_r^2 - \omega^2)^2 + \gamma^2\omega^2}\right](\omega d/c). \tag{14}$$

We emphasize, once again, that the fields radiated by the oscillating dipoles of the thin slab in the forward (transmission) and backward (reflection) directions are identical, and that the crucial difference between the reflected and transmitted beams is the coherent superposition of the (slightly delayed) incident beam, $E^{(i)}\exp(i\omega d/c)$, onto the co-propagating radiated beam in the forward direction. We also remind the reader that all the parameters appearing in Eqs.(13) and (14), namely, $\omega, d, \chi_b, \omega_p, \omega_r, \gamma$, are real-valued and positive. In each equation, the upper sign (+ or −) corresponds to an absorptive (i.e., lossy) medium, whereas the lower sign (− or +) represents an amplifying (i.e., gainy) medium of the thin slab.

**2. Reflection and transmission of a narrow-band light pulse.** Suppose now that the incident $E$-field is a narrow-linewidth pulse of center frequency $\omega_0$ and linewidth $2\Delta\omega$ in the form of

$$\tilde{E}^{(i)}(t) = (2\pi)^{-1}\int_{-\infty}^{\infty} E^{(i)}(\omega)\exp(-i\omega t)\,d\omega. \tag{15}$$

In Appendix A we show that the incident optical energy per unit cross-sectional area is given by

$$\mathcal{E}^{(i)} = Z_0^{-1}\int_{-\infty}^{\infty}\tilde{E}^{(i)2}(t)dt = \frac{1}{2\pi Z_0}\int_{-\infty}^{\infty}|E^{(i)}(\omega)|^2 d\omega. \tag{16}$$

Here $Z_0 = \sqrt{\mu_0/\varepsilon_0}$ is the impedance of free space. Similar expressions as above hold for the reflected and transmitted beams.

If one assumes that the center frequency $\omega_0$ of the incident beam is far from the resonance frequency $\omega_r$ of the material medium, then, for all the incident frequencies in the relevant range $(\omega_0 - \Delta\omega) \le \omega \le (\omega_0 + \Delta\omega)$, one will have $|(\omega_r/\omega)^2 - 1|\omega \gg \gamma$. To gain a better appreciation for the order-of-magnitude estimates used in approximating Eqs.(13) and (14), let $d = 5.0$ Å $= 5\times 10^{-10}$ m, $c = 3\times 10^8$ m/s, $\omega_0 = 3\times 10^{15}$ rad/s (corresponding to a vacuum wavelength $\lambda_0 \cong 0.63$ μm), $\Delta\omega = 10^{13}$ rad/s, $\chi_b = 4.0$, $\omega_r = 6\times 10^{15}$ rad/s, $\gamma = 5\times 10^{14}$ rad/s, and $\omega_p = \sqrt{Ne^2/(\varepsilon_0 m_e)} \cong 10^{16}$ rad/s. (In the latter expression, $N$ is the number-density of the oscillators, $e$ is the charge of an electron, and $m_e$ is the mass of an electron.) With these parameters, the expression on the right-hand side of Eq.(14), evaluated at $\omega = \omega_0$, becomes $(1\mp 0.000513) + i(3 \pm 1.85)(0.005)$. The transmissivity of the slab in the vicinity of $\omega_0$ may now be approximated as follows:

$$T(\omega) = |E^{(t)}/E^{(i)}|^2 \cong \left[1 \mp \frac{\tfrac{1}{2}(\omega_p^2\gamma d/c)}{[(\omega_r/\omega)^2-1]^2\omega^2 + \gamma^2}\right]^2 + \left[1 + \tfrac{1}{2}\chi_b \pm \frac{\tfrac{1}{2}\omega_p^2[(\omega_r/\omega)^2-1]}{[(\omega_r/\omega)^2-1]^2\omega^2 + \gamma^2}\right]^2(\omega d/c)^2$$

$$\cong 1 \mp \frac{\omega_p^2\gamma d/c}{[(\omega_r/\omega)^2-1]^2\omega^2 + \gamma^2} \cong 1 \mp \frac{\omega_p^2\gamma d/c}{[(\omega_r/\omega_0)^2-1]^2\omega_0^2}. \tag{17}$$

Note that the loss (or gain) upon transmission through the slab is proportional to both $\gamma$ and $d$, and that, in any case, this loss (or gain) is fairly small when $\omega_0$ is well below the resonance frequency $\omega_r$. For the parameter set mentioned earlier, Eq.(17) yields $T \cong 1 \mp 0.001$. The exact values for the reflectance $R(\omega) = |E^{(r)}/E^{(i)}|^2$ and transmittance $T(\omega) = |E^{(t)}/E^{(i)}|^2$, obtained from Eqs.(9) and (10) at $\omega = \omega_0$, are $(R = 0.3695\times 10^{-3}, T = 0.9986)$ for a lossy slab and $(R = 0.8556\times 10^{-6}, T = 1.0010)$ for a gainy slab.

Next, we examine the case where the center frequency $\omega_0$ of the incident pulse is at (or in close proximity of) the resonance frequency $\omega_r$ of the material medium. Within the relevant frequency range, $|\omega - \omega_r| \leq \Delta\omega$, a good approximation to the transmissivity of the slab is given by the penultimate expression in Eq.(17), namely,

$$T(\omega) = |E^{(t)}/E^{(i)}|^2 \cong 1 \mp \frac{\omega_p^2 \gamma d/c}{[(\omega_r/\omega)^2 - 1]^2 \omega^2 + \gamma^2}. \tag{18}$$

The loss (or gain) factor is now seen to be confined to a narrow range of frequencies in the vicinity of $\omega_r$, as shown in Fig.2(a). The plot in Fig.2(b) depicts the frequency-dependence of another term that also appears in Eq.(17), but can be shown to be negligible in the limit where $d \ll \lambda_0 = 2\pi c/\omega_0$.

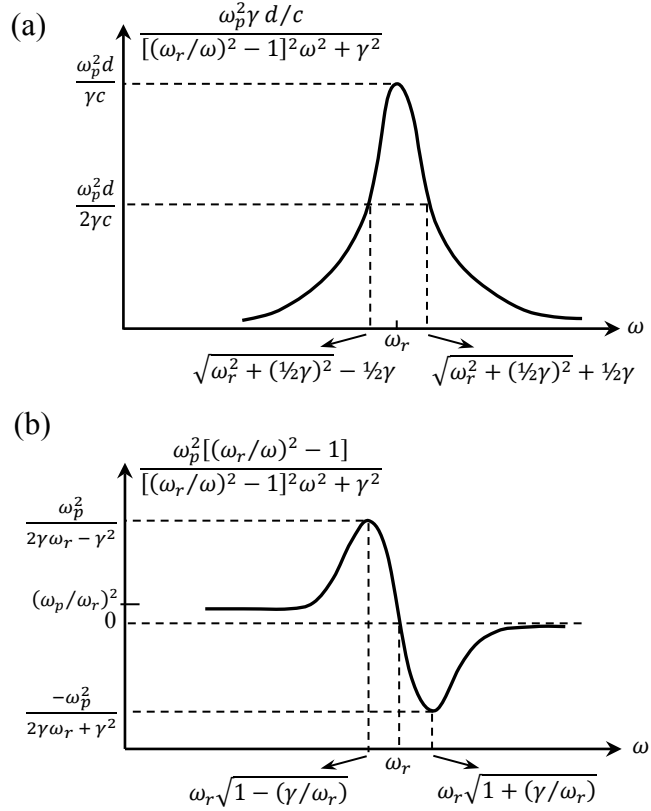

**Fig. 2**. Plots of two functions that appear in Eqs.(13) and (14) versus the incident frequency $\omega$. Both functions peak at or near the resonance frequency $\omega_r$. For $\gamma \ll \omega_r$, the width $\delta\omega$ of the loss (or gain) region in the neighborhood of $\omega_r$ is proportional to $\gamma$. Whereas the function depicted in (a) contributes significantly to the transmissivity of the slab, the contribution of the function shown in (b) is negligible — for sufficiently thin slabs — in a first approximation.

If the linewidth $\Delta\omega$ of the incident pulse were much narrower than the natural linewidth $\delta\omega = \gamma$ of the slab's material, the transmissivity $T$ would vary strongly with detuning $\omega_0 - \omega_r$, as can be inferred from the graph in Fig.2(a). For a *broadband* pulse centered at or near $\omega_0 = \omega_r$, however, the overall loss (or gain) upon transmission will depend on the area under the function depicted in Fig.2(a), which is obtained by evaluating the following integral (see Appendix B):

$$\int_0^\infty \frac{(\omega_p^2 \gamma d/c)\omega^2}{(\omega_r^2 - \omega^2)^2 + \gamma^2 \omega^2} d\omega = \tfrac{1}{2}\pi \omega_p^2 d/c. \tag{19}$$

The integrated loss (or gain) over all frequencies $\omega$ is thus seen to be independent of $\gamma$, as the contributions of $\gamma$ to the height and width of the loss (or gain) function depicted in Fig.2(a) cancel out. Note that the loss (or gain) is primarily associated with the transmitted beam, since the reflected beam is too weak to carry a significant fraction of the incident optical energy. This is a hallmark of stimulated emission, where the excited medium radiates in the forward (i.e., transmission) direction.

It might seem strange at first that the overall transmitted beam is far stronger than the reflected beam, even though the dipoles that are excited within the slab radiate symmetrically in both forward and backward directions. The fundamental physical reason behind this curious behavior is the coherent superposition of the forward-radiated pulse with the intact (albeit delayed) incident pulse after its passage through the slab. The superposition of the two waves in the transmission direction gives rise to constructive interference in the case of gainy slabs, and to destructive interference in the case of lossy slabs. Either way, the electromagnetic energy-density of the transmitted beam, being proportional to the square of the corresponding $E$-field, will have a large cross-term in consequence of interference between the (delayed) incident beam — emerging after passage through the slab — and the forward-radiated beam. The cross-term, which is eminently present in the transmitted pulse, contributes substantially to the transmitted energy, whereas the reflected pulse, being solely the result of backward radiation by the dipoles, does not benefit from a similar boosting effect of interference with a stronger beam.

**3. Conclusions**. Assuming an incident pulse centered at $\omega_r$, with a more-or-less flat spectrum of amplitude $E_0(\omega_r)$, and an spectral width $\sim 2\gamma$, the incident optical energy (per unit area of the slab) will be proportional to $2\gamma|E_0|^2$, while the corresponding energy loss (or gain) will be ½$(\pi\omega_p^2 d/c)|E_0|^2$. The fraction of incident energy that is absorbed (or emitted by stimulation) is thus $\pi\omega_p^2 d/(4\gamma c)$. Denoting the time-averaged magnitude of the incident Poynting vector by $\langle S^{(i)}\rangle$, the rate of energy loss (or gain) per oscillator is given by $B\langle S^{(i)}\rangle$, where $B$ (in units of meter$^2$) is the Einstein coefficient of absorption (or stimulated emission) by individual oscillators. Given that the number of oscillators per unit area of the slab is $Nd$, the rate $NdB\langle S^{(i)}\rangle$ of energy loss (or gain) per unit area must equal $[\pi\omega_p^2 d/(4\gamma c)]\langle S^{(i)}\rangle$. We conclude that $B = \pi Z_0 q^2/(4m_e\gamma)$. [Note that the rate $B\langle S^{(i)}\rangle$ of energy loss (or gain) per oscillator may equivalently be expressed in terms of the incident $E$-field *intensity*, in which case the coefficient $Z_0$ in the above expression of the Einstein $B$ coefficient will move over to $\langle S^{(i)}\rangle$, changing the time-averaged Poynting vector to time-averaged $E$-field intensity.] If an incident light pulse of cross-sectional area $a$, duration $\tau$, frequency $\omega_r$, linewidth $\sim 2\gamma$, and Poynting vector $\langle S^{(i)}\rangle$, happens to have energy $\mathcal{E}^{(i)} = \hbar\omega_r = a\tau\langle S^{(i)}\rangle$, then the absorbed (emitted) energy will be $\tau B\langle S^{(i)}\rangle = (B/a)\hbar\omega_r$. In other words, the probability of single-photon absorption (or stimulated emission) is given by $B/a$.

The rate of spontaneous emission by individual oscillators, known as the Einstein $A$ coefficient, may also be written as $B\langle S^{(i)}\rangle$, provided that $\langle S^{(i)}\rangle$ is interpreted as a superposition of time-averaged Poynting vectors associated with vacuum fluctuations — i.e., the ground state of all the vacuum modes of the electromagnetic field. Given that the mode-density at $\omega_r$ is $\omega_r^2\delta\omega/(\pi^2 c^3)$, that each vacuum mode has energy ½$\hbar\omega_r$, and that this energy propagates with the speed of light $c$ in vacuum, we find $A/B = \langle S_{\text{vac}}\rangle = ½\hbar\omega_r^3\delta\omega/(\pi^2 c^2)$. This ratio of $A/B$ is off by a factor of 2 compared to that obtained by Einstein [3]. The discrepancy is resolved when the effect of radiation resistance on the spontaneous emission process is taken into account [9].

## Appendix A

To prove the identity in Eq.(16), we substitute for $\tilde{E}^{(i)}(t)$ from Eq.(15), as follows:

$$\mathcal{E}^{(i)} = Z_0^{-1} \int_{-\infty}^{\infty} \tilde{E}^{(i)^2}(t) dt$$

$$= \frac{1}{(2\pi)^2 Z_0} \iint_{-\infty}^{\infty} E^{(i)}(\omega) E^{(i)}(\omega') \int_{-\infty}^{\infty} \exp[-i(\omega + \omega')t] dt \, d\omega d\omega'$$

$$= \frac{1}{2\pi Z_0} \iint_{-\infty}^{\infty} E^{(i)}(\omega) E^{(i)}(\omega') \delta(\omega + \omega') d\omega d\omega'$$

$$= \frac{1}{2\pi Z_0} \int_{-\infty}^{\infty} E^{(i)}(\omega) E^{(i)}(-\omega) d\omega = \frac{1}{2\pi Z_0} \int_{-\infty}^{\infty} |E^{(i)}(\omega)|^2 d\omega.$$

In the above derivation we have used the fact that $\int_{-\infty}^{\infty} \exp(-i\omega t) dt = 2\pi\delta(\omega)$, and also that $E^{(i)}(\omega)$, being the Fourier transform of the real-valued function $\tilde{E}^{(i)}(t)$, is Hermitian, that is, $E^{(i)}(-\omega) = E^{(i)^*}(\omega)$.

## Appendix B

To evaluate the integral in Eq.(19), note that the denominator is a 4$^{th}$ order polynomial in $\omega$, whose roots are readily found to be

$$(\omega_r^2 - \omega^2)^2 + \gamma^2 \omega^2 = 0 \quad \to \quad \omega^2 \pm i\gamma\omega - \omega_r^2 = 0 \quad \to \quad \omega_{1,2,3,4} = \tfrac{1}{2}\gamma\left[\pm\sqrt{(2\omega_r/\gamma)^2 - 1} \pm i\right].$$

Here we shall assume that $\gamma < 2\omega_r$. The roots in the upper-half of the complex plane are denoted by $\omega_1$ and $\omega_2$, while those in the lower-half are $\omega_3 = \omega_1^*$ and $\omega_4 = \omega_2^*$. The integral on the $\omega$-axis is evaluated around a closed contour in the upper half of the complex-plane. Invoking Cauchy's theorem and using the residues of the integrand at $\omega_1$ and $\omega_2$, we find

$$\int_0^\infty \frac{(\omega_p^2 \gamma d/c)\omega^2}{(\omega_r^2 - \omega^2)^2 + \gamma^2 \omega^2} d\omega = \tfrac{1}{2}(\omega_p^2 \gamma d/c) \int_{-\infty}^\infty \frac{\omega^2}{(\omega-\omega_1)(\omega-\omega_2)(\omega-\omega_1^*)(\omega-\omega_2^*)} d\omega$$

$$= i\pi(\omega_p^2 \gamma d/c) \left[\frac{\omega_1^2}{(\omega_1-\omega_2)(\omega_1-\omega_1^*)(\omega_1-\omega_2^*)} + \frac{\omega_2^2}{(\omega_2-\omega_1)(\omega_2-\omega_1^*)(\omega_2-\omega_2^*)}\right]$$

$$= \frac{i\pi(\omega_p^2 \gamma d/c)}{\omega_1 - \omega_2} \left[\frac{\omega_1^2}{2i\,\text{Imag}(\omega_1)(\omega_1-\omega_2^*)} + \frac{\omega_2^2}{2i\,\text{Imag}(\omega_2)(\omega_1^*-\omega_2)}\right]$$

$$= \frac{\pi(\omega_p^2 \gamma d/c)}{\gamma^2 \sqrt{(2\omega_r/\gamma)^2 - 1}} \left(\frac{\omega_1^2}{2\omega_1} - \frac{\omega_2^2}{2\omega_2}\right) = \tfrac{1}{2}\pi \omega_p^2 \, d/c.$$

An alternative evaluation of the integral using the *Table of Integrals*[†] leads to the same result:

---

[†]I. S. Gradshteyn and I. M. Ryzhik, *Table of Integrals, Series, and Products*, 7$^{th}$ edition, Academic Press, New York (2007); see equation **3.252**-12:

$$\int_0^\infty \frac{x^{\mu-1}}{x^2 + 2ax \cos t + a^2} dx = -\frac{\pi a^{\mu-2} \sin[(\mu-1)t]}{\sin(t)\sin(\mu\pi)}; \quad a > 0, \quad 0 < |t| < \pi, \quad 0 < \text{Re}(\mu) < 2.$$

Here, $a = \omega_r^2$, $\mu = 3/2$, $t \cong \pi - (\gamma/\omega_r) \quad \to \quad \cos t = -1 + \tfrac{1}{2}(\gamma/\omega_r)^2$.

$$\int_0^\infty \frac{(\omega_p^2 \gamma d/c)\omega^2}{(\omega_r^2 - \omega^2)^2 + \gamma^2 \omega^2} \mathrm{d}\omega = (\omega_p^2 \gamma\, d/c) \int_0^\infty \frac{\omega^2}{\omega^4 - 2\omega_r^2[1 - \tfrac{1}{2}(\gamma/\omega_r)^2]\omega^2 + \omega_r^4} \mathrm{d}\omega$$

$$= \tfrac{1}{2}(\omega_p^2 \gamma\, d/c) \int_0^\infty \frac{\sqrt{x}}{x^2 - 2\omega_r^2[1 - \tfrac{1}{2}(\gamma/\omega_r)^2]x + \omega_r^4} \mathrm{d}x$$

$$\cong \tfrac{1}{2}(\omega_p^2 \gamma\, d/c)[-(\pi/\omega_r)(\omega_r/\gamma)(-1)\cos(\gamma/2\omega_r)] \cong \tfrac{1}{2}\pi \omega_p^2\, d/c.$$